\documentclass[twocolumn,pra,showpacs,showkeys]{revtex4}   % use preprint or twocolumn
\usepackage{graphicx}
\usepackage{amsmath}
\usepackage{amssymb}
\usepackage{epstopdf}
\usepackage{dcolumn}% Align table columns on decimal point
\usepackage{bm}% bold math

\newcommand*{\etal}{\emph{et al.}~}

\newcommand*{\lp}{\left(}
\newcommand*{\rp}{\right)}

\begin{document}

\title{Absolute and ratio measurements of the polarizability of Na, K, and Rb with an atom interferometer}

\author{William F. Holmgren}
\author{Melissa C. Revelle}
\author{Vincent P. A. Lonij}
\author{Alexander D. Cronin}%
 \email{cronin@physics.arizona.edu}
\affiliation{%
Department of Physics, University of Arizona, Tucson, AZ 85721
}%

\date{\today}        

\begin{abstract}
We measured the ground state electric dipole polarizability of sodium, potassium, and rubidium using a Mach-Zehnder atom interferometer with an electric field gradient. We find $\alpha_{\textrm{Na}}=24.11(2)_\textrm{stat}(18)_\textrm{sys} \times 10^{-24} \textrm{ cm}^3$, $\alpha_{\textrm{K}}=43.06(14)(33)$, and $\alpha_{\textrm{Rb}}=47.24(12)(42)$. Since these measurements were all performed in the same apparatus and subject to the same systematic errors we can present polarizability ratios with 0.3\% precision. We find $\alpha_{\textrm{Rb}}/\alpha_{\textrm{Na}}=1.959(5)$, $\alpha_{\textrm{K}}/\alpha_{\textrm{Na}}=1.786(6)$, and $\alpha_{\textrm{Rb}}/\alpha_{\textrm{K}}=1.097(5)$. We combine our ratio measurements with the higher precision measurement of sodium polarizability by Ekstrom \etal [Phys.~Rev.~A 51, 3883 (1995)] to find $\alpha_{\textrm{K}}=43.06(21)$ and $\alpha_{\textrm{Rb}}=47.24(21)$.
\end{abstract}



\maketitle

\section{Introduction}

Precision measurements of polarizability serve as benchmark tests for methods used to model atoms and molecules \cite{Gou05,Der99}.   Accurate calculations of van der Waals interactions, state lifetimes, branching ratios, indices of refraction and polarizabilities all rely on sophisticated many-body theories with relativistic corrections, and all of these quantities can be expressed in terms of atomic dipole matrix elements. Polarizability measurements, such as the ones presented here, are some of the best ways to test these calculations.

Over 35 years ago, Molof \etal \cite{Mol74} measured alkali metal and metastable noble gas polarizabilities with an uncertainty of 2\% using beam deflection and the E-H gradient balance technique. More recently, atom interferometers were used to measure the polarizability of lithium \cite{Mif06} and sodium  \cite{Eks95} with an uncertainty of 0.7\% and 0.35\%, respectively. Near-field molecule interferometry was used to measure the polarizability of $\textrm{C}_{60}$ and $\textrm{C}_{70}$ with 6\% uncertainty \cite{Ber07Arndt}, and guided BEC interferometry was used to measure the dynamic polarizability of rubidium with 7\% uncertainty \cite{Dei08}.   A fountain experiment was used to measure the polarizability of cesium with 0.14\% precision \cite{Ami03}.  The measurements of potassium and rubidium polarizability made by Molof \etal remained the most precise until now. 

In this paper we present absolute and ratio measurements of the ground state electric dipole polarizability of sodium, potassium, and rubidium using a Mach-Zehnder atom interferometer with an electric field gradient. The uncertainty of each absolute measurement is less than 1.0\% and the precision of each ratio measurement is 0.3\%. Our interferometer is constructed with nanogratings that diffract all types of atoms and molecules and enable us to measure the polarizabilities of different atomic species in the same apparatus. The systematic errors are nearly the same for the different atomic species and cancel when calculating polarizability ratios. Finally, we combine our polarizability ratios with the absolute measurement of sodium polarizability by Ekstrom \etal \cite{Eks95} to provide measurements of potassium and rubidium polarizabilities with 0.5\% uncertainty. 

A unique feature of this work compared to references \cite{Mif06, Eks95} is that we use an electric field gradient region rather than a septum electrode. In addition, we use a less collimated beam to increase flux and reduce systematic error caused by velocity selective detection of atoms in the interferometer.

\section{Apparatus}\label{Apparatus}

\begin{figure}[b]
\includegraphics{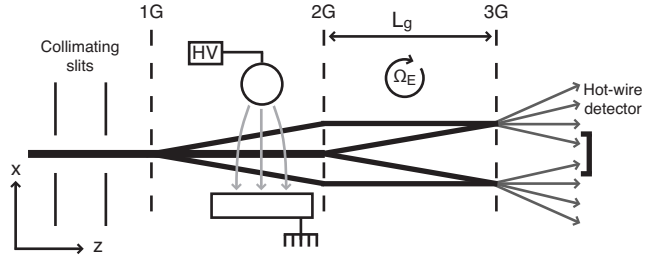}
\caption{\label{IFMgradEfig}Nanogratings 1G, 2G, and 3G form multiple Mach-Zehnder interferometers (two are shown). An atom passing through the interaction region acquires a phase $\phi_1$, $\phi_0$, and $\phi_{-1}$ along each path. The third grating acts as a mask for the 100 nm period interference fringes and also diffracts the interferometer output. The hot-wire detector is centered on the 0th-order path. The distance between two gratings is $L_g=940\textrm{ mm}$. The vertical (transverse) scale is exaggerated $10^4$ times. The Earth rotation rate $\Omega_E$ modifies the measured phase shift.}
\end{figure}

Our apparatus is described in detail elsewhere \cite{Cro09, Ber97}. In brief, we use three 100 nm period nanogratings to diffract a supersonic beam of sodium, potassium, or rubidium atoms and form multiple Mach-Zehnder interferometers (see Fig.~\ref{IFMgradEfig}). An atom diffracted by the first and second gratings may be found with a sinusoidal probability distribution at the plane of the third grating. The third grating acts as a mask of this interference pattern and also diffracts the interferometer output. We measure the flux as a function of grating position to determine the phase and contrast of the fringe pattern. We detect $10^5$ atoms/sec with a typical contrast of 30\% using a hot-wire detector 0.5 m beyond the third grating. 

We measure the output of the two interferometers formed by first order diffraction from the first and second nanogratings (see Fig.~\ref{IFMgradEfig}). Although other interferometers are present, they do not contribute to the measured phase shift because they either are not white-light interferometers, have fringes with a periodicity different than that of the third grating, or are simply not incident upon the detector.   The interferometers formed by 2nd order diffraction from the first grating \cite{Per06} contribute less than 1\% of the detected signal and cause an error in our polarizability measurements of less than 0.01\%.

Before the second grating the path separation in the interferometer is 
\begin{align}
\label{sepeqn}
s=\frac{\lambda_{dB}}{d_g}z=\frac{h}{mvd_g}z
\end{align}
where $\lambda_{dB}=h/mv$ is the de Broglie wavelength of an atom with mass $m$ and velocity $v$, $d_g$ is the grating period, and $z$ is the propagation distance from the first grating. We adjust the beam velocity for each atomic species such that $s\approx50$ microns in the interaction region, where the beam width of each diffraction order is approximately 80 microns. We designed the beam parameters to be similar 
for each atomic species in order to minimize systematic errors in measurements of polarizability ratios. 

As in previous work \cite{Eks95, Mif06, Ber07Arndt}, we place an interaction region between the first and second gratings to induce a differential phase shift in the interferometer. The phase shift is proportional to the atomic polarizability. Unlike references \cite{Mif06,Eks95}, we use an electric field gradient region rather than a septum electrode  as an interaction region.   We use an electric field gradient because the septum electrode would require fully separated diffraction orders and this is more difficult with heavier atoms such as potassium and rubidium.

\begin{figure}
\includegraphics{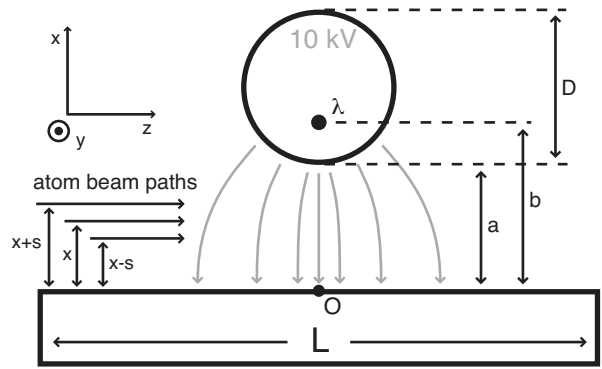}
\caption{\label{GradEgeometry}Cross section of the interaction region (not to scale). The high voltage electrode of diameter $D=12.66 \textrm{ mm}$ is fixed at a distance $a=1.998 \textrm{ mm}$ from the ground plane by precision spacers (not shown). The effective line charge $\lambda$ is located a distance $b$ from the ground plane, as discussed in the text. The ground plane is of length $L=90 \textrm{ mm}$. The high voltage electrode and ground plane are 50 mm long in the y direction, while the beam height is only 1 mm. The 0th order beam is a distance $x$ from the ground plane and the $\pm1$st order beams are a distance $x\pm s$ from the ground plane. Electric field lines are shown in gray. The beam propagates along the $z$ axis. $O$ is the origin for the electric field calculations.}
\end{figure}

The geometry of our interaction region is depicted in Fig.~\ref{GradEgeometry}. The interaction region consists of a cylindrical electrode and a grounded plane.  This geometry is the familiar ``two-wire" configuration \cite{Ram56} rotated by 90 degrees so that the height of the cylinder electrode is perpendicular, rather than parallel, to the beam paths. Our electrode orientation yields a relatively small fringe displacement (200 nm) compared to the standard electrode orientation for Stark deflections (200 $\mu$m)  \cite{Mol74, Hal74, Tar93, Tik01, Sch07, Sch08}, but the sensitivity of atom interferometry allows us to make precise measurements of such small deflections. Two advantages of our electrode orientation are that the phase shift is homogeneous  across the height of the atom beam and there are no fringing fields entering and exiting the interaction region.

We apply a voltage of 0-12 kV to the cylindrical electrode to create the electric field gradient. Our electrode geometry is easily analyzed via the method of images \cite{Wan86}. The boundary conditions of our geometry, with cylindrical symmetry and an infinite ground plane, correspond exactly to the geometry in which an infinitely long line charge $\lambda$ is fixed a distance $b$ from the ground plane. The equipotential surfaces are circles of increasing radius centered at an increasing distance from the ground plane. We identify one of these equipotential surfaces as our electrode at a voltage $V$ with radius $R$ and located a distance $a$ from the ground plane to determine the corresponding effective line charge $\lambda$ and its position $b$:
\begin{align}
\lambda&=2\pi\epsilon_0V\ln^{-1}\lp\frac{a+R+b}{a+R-b}\rp\\
b&=a\sqrt{1+2R/a}.
\end{align}
The resulting electric field is given by
\begin{align}
\bm{E}_{(x,z)} = \frac{\lambda}{\pi\epsilon_0} \bigg[ \left( \frac{x-b}{(x-b)^2+z^2}  - \frac{x+b}{(x+b)^2+z^2}\right) \hat{x} \nonumber\\
+ \left(  \frac{z}{(x-b)^2+z^2}- \frac{z}{(x+b)^2+z^2}\right) \hat{z} \bigg].
\end{align}

The potential energy of an atom in an electric field is given by the Stark shift $U_{\textrm{Stark}}=-\frac{1}{2}\alpha E^2$. We use the WKB approximation to find the phase $\phi_\alpha(x,v)$ acquired by an atom along a path a distance $x$ from the ground plane with velocity $v$ and polarizability $\alpha$:
\begin{align}
\phi_\alpha(x,v)& = \frac{\alpha}{2\hbar v} \int^{\infty}_{-\infty} E^2_{(x,z)}\, dz.
\end{align}
For our atom beam $U_{\textrm{Stark}}\approx10^{-9}\textrm{ eV}$ and $U_{\textrm{kinetic}}\approx 0.1\textrm{ eV}$, so the WKB approximation is valid. The integral of $E^2$ along the path of the atom may be performed using complex analysis and yields an acquired phase of
\begin{align}
\phi_\alpha(x,v) &= \frac{\lambda^2\alpha}{\pi\epsilon^2_0\hbar v}\lp\frac{b}{b^2-x^2}\rp .
\end{align}  We induce a polarizability phase $\phi_\alpha$ of up to 2500 rad along one path.

We will now discuss how the phase and contrast of the measured fringe pattern depends on the polarizability phase $\phi_\alpha(x,v)$. First, we define the phase difference between the paths of the two detected interferometers:
\begin{align}
\phi_{\alpha,1}(x,v)&=\phi_\alpha(x+s,v)-\phi_\alpha(x,v) \nonumber \\
\\
\phi_{\alpha,-1}(x,v)&=\phi_\alpha(x,v)-\phi_\alpha(x-s,v).\nonumber
\end{align}
We studied phase differences $\phi_{\alpha,1}$ of up to 18 rad.
Next, we perform an incoherent sum of the fringe patterns formed by atoms of multiple velocities traversing multiple interferometers. The resulting fringe pattern is described by
\begin{align}
\label{incsum}
C_{\textrm{sum}}(x)e^{i\phi_\textrm{sum}(x)}=C_0e^{i\phi_0}\sum_{j=-1,1}P_j\int^\infty_0 P(v)e^{i\phi_{\alpha,j}(x,v)}dv
\end{align}
where $C_{\textrm{sum}}$ is the real-valued contrast of the fringe pattern, $\phi_{\textrm{sum}}$ is the phase of the fringe pattern, $C_0$ and $\phi_0$ are the initial contrast and phase of the interferometer, $j$ denotes the interferometer number (upper or lower diamond in Fig.~\ref{IFMgradEfig}), $P_j=0.5$ is the probability of an atom being found in interferometer $j$, and $P(v)$ is the velocity distribution of the beam. In our experiment, the phase shift $\phi_{\textrm{sum}}$ is reduced by as much as 4\% by performing the sum described in eqn.~(\ref{incsum}) compared to a simple weighted average of phases, and the contrast is reduced by more than 50\%.

The Sagnac phase must also be accounted for in our experiment and modifies eqn.~(\ref{incsum}) \cite{Len97,Jac08}. Because the Sagnac phase is dispersive, ignoring it would lead to an error in polarizability of up to 1\%. The Sagnac phase in our interferometer is given by
\begin{align}
\label{phiSagofv}
\phi_{\textrm{Sag}}(v)=4\pi{L_{g}}^2\Omega/d_gv
\end{align}
where $L_{g}$ is the distance between adjacent nanogratings and $\Omega$ is the rotation rate of the Earth projected into the plane of the interferometer. At our latitude, the Sagnac phase is as much as 4.8 rad for our rubidium beam. The reference phase, $\phi_\textrm{ref}$, and contrast, $C_\textrm{ref}$, of the interferometer are determined by the Sagnac phase in the absence of an electric field: 
\begin{align}
\label{phiSaginc}
C_{\textrm{ref}}e^{i\phi_\textrm{ref}}=C_0e^{i\phi_0}\sum_{i=-1,1}P_i\int^\infty_0 P(v)e^{i\phi_{\textrm{Sag}}(v)}dv.
\end{align}
We find the total phase and contrast of the interferometer in the presence of an electric field by adding the Sagnac phase to the polarizability phase shift before conducting the incoherent sum shown in eqn.~(\ref{incsum}). This procedure yields
\begin{multline}
\label{incsumPolSag}
C_{\textrm{total}}(x)e^{i\phi_{\textrm{total}}(x)}=\\
C_0e^{i\phi_0}\sum_{i=-1,1}P_i\int^\infty_0 P(v)e^{i[\phi_{\alpha,i}(x,v)+\phi_{\textrm{Sag}}(v)]}dv.
\end{multline}
Finally, the measured phase shift and contrast are 
\begin{align}
\label{phimeasSag}
\phi_{\textrm{measured}}(x)=\phi_{\textrm{total}}(x)-\phi_{\textrm{ref}}
\end{align}
\begin{align}
\label{conmeasSag}
C_{\textrm{measured}}(x)=C_{\textrm{total}}/C_{\textrm{ref}}.
\end{align}

As an alternative point of view, we may describe the measured phase shift in terms of a classical electrostatic force on the individual atomic dipoles instead of the quantum mechanical phases acquired by an atom in the electric field. In the classical mechanics picture, a neutral atom in an electric field experiences a force $\bm{F}=-\bm{\nabla} U_{\textrm{Stark}}=\alpha \bm{E} \bm{\nabla} \bm{E}$. The deflection of the interferometer paths will cause the same displacement of the observed fringes as the phase shift analysis discussed above.

\section{Velocity Measurement}\label{VelocitySection}

The velocity determines both the amount of time an atom interacts with the electric field and the spatial separation $s$ of the paths inside the electric field gradient. Therefore, an accurate determination of the beam velocity and the velocity distribution is essential for a precise polarizability measurement.

\begin{figure}[t]
\includegraphics{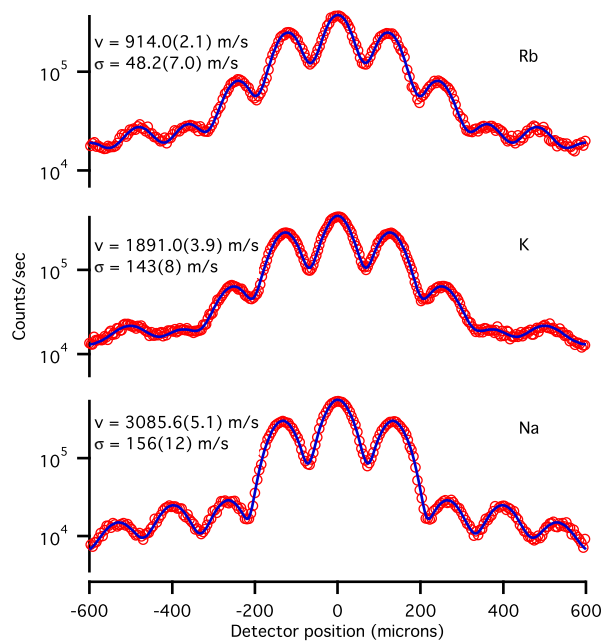}
\caption{\label{Diffraction}(Color online) Diffraction of Rb, K, and Na atoms from the same nanograting. Best-fit flow velocity $v_0$ and velocity distribution $\sigma_v$ with statistical errors are shown. As discussed in the text, the velocity distribution is found from contrast loss measurements.}
\end{figure}

We determine the velocity of the atom beam by analyzing the far-field diffraction pattern from the first grating. The velocity distribution of the beam is modeled by  
\begin{align}
P(v)dv = A v^3 \exp\lp-\lp v-v_0\rp^2/2\sigma_v^2\rp dv
\end{align}
where $v$ is the velocity, $v_0$ is the flow velocity, $\sigma_v$ describes the velocity distribution, and $A$ is a normalization factor \cite{Hab85}. In the limit of a supersonic beam, $v_0/\sigma_v\gg1$, the normalization factor can be written as $A=(\sqrt{2\pi}v_0\sigma_v({v_0}^2+3{\sigma_v}^2))^{-1}$. The location of the $n$th diffraction order at the detector plane is given by
\begin{align}
\label{sepeqnN}
x_n=\frac{\lambda_{dB}}{d_g}nz_{det}=\frac{hn}{mvd_g}z_{det}
\end{align}
where the propagation distance $z$ is equal to the distance from the first grating to the detector, $z_{det}$. We use $m=m_{\textrm{avg}}$, the average mass of the atomic species, rather than calculating and adding the diffraction patterns for each isotope. A reanalysis of a subset of our data shows that this approximation yields a small difference in velocity ($<0.02\%$) and polarizability ($<0.05\%$) when isotopes are accounted for. Next, we rearrange eqn.~(\ref{sepeqnN}) to find 
\begin{align}
v(x_n)=z_{det}hn/md_gx_n
\end{align}
and use this to transform $P(v)dv$ to $P(x)dx$. Finally, we sum over all diffraction orders, each weighted by $c_n$, and add the 0th order peak to obtain the diffraction pattern for an infinitesimally thin beam and detector: 
\begin{widetext}
\begin{align}
\label{pofx}
P(x)dx=\left\{ c_0\delta(x-0)+ \sum_{n\neq0} c_n A\lp\frac{z_{det}hn}{md_g}\rp^4 x^5 \exp\left[-\lp \frac{z_{det}hn}{md_gx}-v_0\rp^2/2\sigma_v^2\right] \right\} dx.
\end{align}
\end{widetext}
The observed diffraction pattern (see Fig.~\ref{Diffraction}) is a convolution of the spatial probability distribution given by eqn.~(\ref{pofx}) with the collimated beam and detector shapes. Two narrow collimating slits of width 20 and 10 microns separated by 890 mm determine the beam shape. We model the detector wire as a square aperture with width 70 microns. We fit the observed diffraction pattern to the convolution described above to find the flow velocity, $v_0$. With four diffraction scans, we can determine $v_0$ with a statistical precision of 0.1\%. 

The diffraction orders are sufficiently close together, the beam sufficiently broad, and the detector sufficiently thick that we cannot use diffraction data alone to determine the velocity distribution $\sigma_v$ with enough precision for the polarizability measurements. Instead, as discussed later, we find the velocity distribution parameter $\sigma_v$ from the contrast loss measurements. We then fix $\sigma_v$ when fitting the diffraction patterns to find the final flow velocity $v_0$.

\section{Phase and Contrast Measurement} \label{PhaseSection}

After recording several diffraction scans to measure the flow velocity, we center the detector on the 0th-order diffraction peak, replace the narrow collimating slits with wider ones (35 and 45 microns), and insert the second and third gratings into the beam line to form the interferometer. We use a wider beam for our interferometer than Ekstrom \etal \cite{Eks95} for two reasons. First, wide collimating slits allow more flux to reach the detector. Second, wide slits minimize the velocity selective detection of interference fringes caused by the dispersive nature of the nanogratings. We calculate that the flow velocity of the atoms detected from the interferometers when the detector wire is centered on the beam is about 0.25\% faster than the flow velocity of the entire beam. We use the adjusted flow velocity when determining the polarizability, yielding a 0.5\% correction to the polarizability. The correction to the velocity distribution parameter $\sigma_v$ is negligible.   If we had used small slits with the detector on the centerline, this correction and the uncertainty in this correction would have been three times larger.

\begin{figure}
\includegraphics{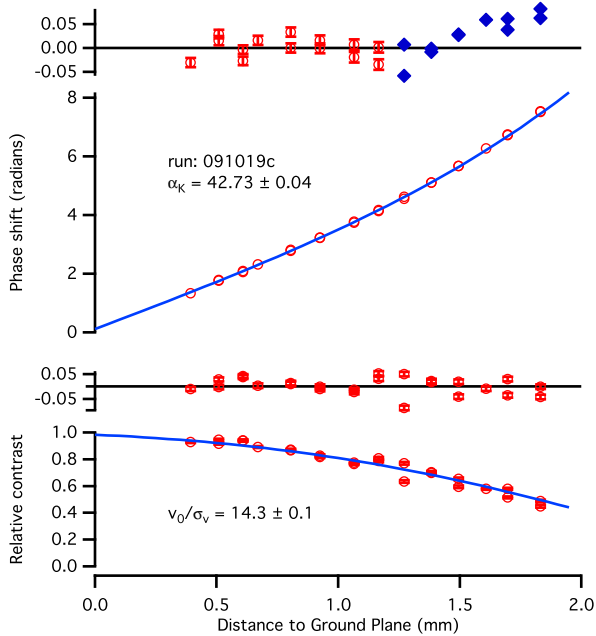}
\caption{\label{PhaseFit}(Color online) Phase shift and relative contrast vs.~electrode position $x$. The best-fit polarizability and the statistical error for one data set is shown. We only fit the phase shift measurements with relative contrast greater than 75\%. Residuals for the fit data points (circles) are shown with error bars. For reference, residuals for the unfit data points (filled diamonds) are also shown. The contrast loss determines $v_0/\sigma_v$.}
\end{figure}

Next, we calibrate the position of the interaction region by eclipsing the beam with the cylindrical electrode and then moving the interaction region out of the beam path as we record the average flux through the interferometer and the position of the interaction region. We use the position at which the flux is 50\% of the maximum to locate the center of the beam a distance $a$ from the ground plane. We move the interaction region then across the beam in steps of 100 $\mu$m and measure the phase shift (eqn.~(\ref{phimeasSag})) and contrast loss (eqn.~(\ref{conmeasSag})) at each position. Figure \ref{PhaseFit} shows the measured phase shift and contrast loss for a typical data set.

We determine the flow velocity, velocity distribution, and polarizability from the diffraction, contrast loss, and phase shift data, respectively. In Section \ref{VelocitySection} we discussed how we find the flow velocity $v_0$. In Section \ref{Apparatus} we discussed how the contrast of the measured fringe pattern is reduced by performing an incoherent sum of the fringes formed by atoms of multiple velocities. We fit the contrast loss data to determine $v_0/\sigma_v$ with an uncertainty of 10\%. The primary source of error in this measurement of $\sigma_v$ comes from vibration-induced fluctuations in the reference contrast. We then refit the diffraction data holding $\sigma_v$ fixed to find the best-fit flow velocity $v_0$. This procedure yields a small correction to $v_0$ of $<0.2\%$. Finally, we use $v_0$ and $\sigma_v$ as inputs to the polarizability fit of the phase data. We exclude data points in which the relative contrast is less than 75\% to minimize the uncertainty in the polarizability due to uncertainty in $\sigma_v$.

After fitting all data, we apply small corrections to the polarizability due to beam thickness and isotope ratios. To account for beam thickness, we modify eqn.~(\ref{incsumPolSag}) to include a sum over beam width. The correction to the polarizability due to beam thickness is +0.04(2)\% for each atomic species. To account for isotope ratios we modify eqns.~(\ref{incsumPolSag}) and (\ref{pofx}) to include weighted sums over isotopes. The correction to the polarizability from taking into account the isotope ratios is +0.04\% for $\alpha_\textrm{K}$ and +0.02\% for $\alpha_\textrm{Rb}$.

\begin{figure}
\includegraphics{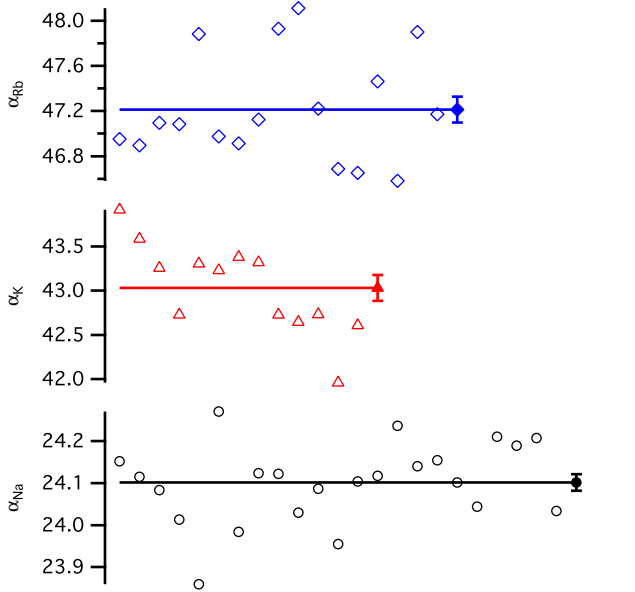}
\caption{\label{Results}(Color online) Multiple measurements of the polarizability of sodium (circles), potassium (triangles), and rubidium (diamonds). The mean polarizabilities are denoted by a filled markers and lines. The error bars represent the standard error of the mean. Units are $10^{-24}\textrm{ cm}^3$.  Final results are shown in Table \ref{ResultsTable}.}
\end{figure}

The result of each data set is shown in Fig.~\ref{Results}. Each point on the plot represents one hour of data.  We report the mean polarizability from all of our  data in Table \ref{ResultsTable}. The reported statistical error is the standard error of the mean and is dominated by the reproducibility of the experiment rather than the statistical phase error of a typical data set. The systematic errors are discussed later. 

\begin{figure}
\includegraphics{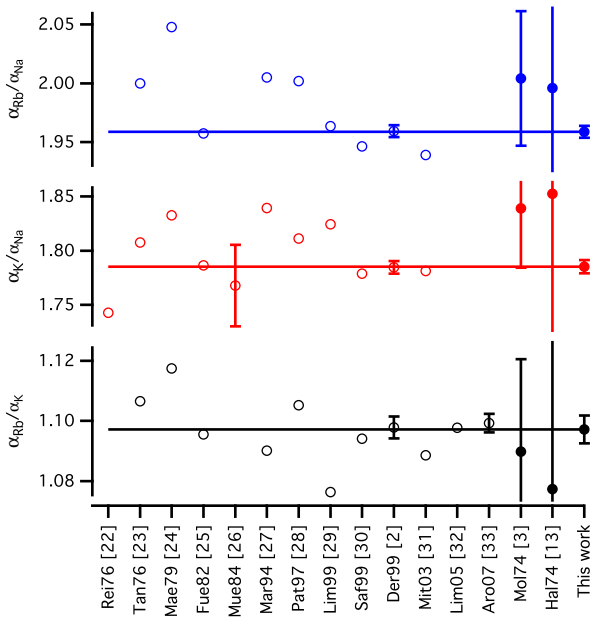}
\caption{\label{PreviousWorkPlot}(Color online) Previously calculated (unfilled) and measured (filled) alkali polarizability ratios. References are denoted by the abbreviated first author's name, the year, and the reference number.}
\end{figure}

Since we performed all measurements in the same apparatus under similar beam conditions and without changing any parameters that contribute to systematic error in the polarizability, we can report polarizability ratios with uncertainties dominated by the statistical precision of our measurements. We show our measured polarizability ratios in Table \ref{RatiosTable}. Figure \ref{PreviousWorkPlot} shows a summary of measurements \cite{Mol74,Hal74} and calculations \cite{Rei76, Tan76, Mae79, Fue82, Mul84, Mar94, Pat97, Lim99, Saf99, Der99, Mit03, Lim05atoms, Aro07} of the polarizability ratios of sodium, potassium, and rubidium, including this work. We added the reported uncertainties for each atom in quadrature to calculate the uncertainty in polarizability ratios for previous work. If the reported uncertainties have systematic errors that would have canceled in ratio measurements, then this calculation will lead to an overestimate of the ratio uncertainties.

\begin{table}
\caption{\label{ResultsTable}Measured absolute and recommended atomic polarizabilities in units of $10^{-24}\textrm{ cm}^3$.  Recommended polarizability measurements are based on our ratio measurements (see Table \ref{RatiosTable}) combined with the sodium polarizability measurement from reference \cite{Eks95}.}
\begin{center}
\begin{tabular}{c c c c}
\hline\hline
& $\alpha_\textrm{abs}$(stat.)(sys.)  & { $\textrm{  }$ $\textrm{  }$ } & $\alpha_{\textrm{rec}}$(tot.)\\
\hline
Na & 24.11(2)(18) & & 24.11(8)\\
K & 43.06(14)(33) & & 43.06(21)\\
Rb & 47.24(12)(42) &  &47.24(21)\\
\hline
\end{tabular}
\end{center}
\end{table}

\begin{table}
\caption{\label{RatiosTable}Measured atomic polarizability ratios with statistical uncertainties. Also included are several polarizability ratios from \emph{ab initio} and semi-empirical calculations. See Fig.~\ref{PreviousWorkPlot} for more previous calculations and measurements of polarizability ratios.}
\begin{center}
\begin{tabular}{c c c c c}
\hline\hline
 &  $\alpha_\textrm{ratio}$(Stat. Unc.) \\
Atoms & This work & Ref \cite{Der99} & Ref \cite{Saf99} & Ref \cite{Mit03} \\
\hline
Rb/Na & 1.959(5) & 1.959(5) & 1.946 & 1.939 \\
K/Na    & 1.786(6) & 1.785(6) & 1.779 & 1.781 \\
Rb/K    & 1.097(5) & 1.098(5) & 1.094 & 1.089 \\
\hline\hline
\end{tabular}
\end{center}
\end{table}

We calculate the highest-precision, recommended measurements of potassium and rubidium polarizability by combining our polarizability ratio measurements with the sodium polarizability measurement by Ekstrom \etal \cite{Eks95}.  To calculate the total uncertainty of the recommended polarizabilities of potassium and rubidium we add the total uncertainty of the Ekstrom \etal sodium measurement in quadrature with the statistical uncertainty of our appropriate polarizability ratio.  Our recommended polarizability values and their total uncertainties are shown in Table \ref{ResultsTable}. Given the 0.8\% uncertainty of our direct measurement of $\alpha_{\textrm{Na}}$, the agreement between our measurement and that of Ekstrom \etal at the level of 0.04\% is coincidental.

Table \ref{ErrorTable} shows a summary of the error budget. Most of the highly significant parameters in the error budget are related to the flow velocity $v_0$ or velocity distribution parameter $\sigma_v$. The most significant parameter in the error budget is the distance from the first grating to the detector, $z_{det}$, due to its effect on our measurement of $v_0$. The details of the beam shape modify the best-fit flow velocity as well. We measure the displacement of our detector translation stage using a Heidenhain MT-2571 length gauge with a linear encoder and fractional uncertainty of 0.02\%. If the detector translation along the x axis is not perpendicular to the beam path along the z axis then we would also report an incorrect velocity. We previously discussed how the velocity selective detection of interfering atoms modifies $v_0$ and adds uncertainty in the polarizability. The effect of the velocity distribution on the measured phase becomes larger as the phase shift increases and the contrast decreases. Therefore, to minimize the uncertainty due to the velocity distribution, we ignore phase data points for which the relative contrast is less than 75\%. This procedure yields an uncertainty of 0.20\% in the polarizability for a 10\% uncertainty in $\sigma_v$. Uncertainty in the distance from the first grating to the interaction region, $z_{int}$, causes uncertainty in the diffracted path separation $s$ in the interaction region. Uncertainty in the electrode spacing $a$, radius $R$, and applied voltage $V$ cause uncertainty in the strength of the electric field. Uncertainty in the electrode orientation about the x, y, and z axes yields a small uncertainty in the polarizability, as well.

The possibility of a small fraction of molecules in the beam contributes an additional source of error. The diffraction scans for the conditions under which we run the interferometer do not have sufficient resolution to determine the molecule fraction of the beam. By reducing the velocity of the beam, and thus increasing the diffraction angle, we found that molecules contribute less than 1\% of the flux. To calculate the corresponding uncertainty in our polarizability measurements we include a sum over two additional molecule interferometers in eqn.~(\ref{incsumPolSag}). We use the molecular polarizabilities measured by Tarnovsky \etal \cite{Tar93} in our calculations to find that the uncertainty in atomic polarizabilities due to the presence of molecules is less than 0.10\%.

An additional source of error comes from the possible tilt of the entire interferometer board with respect to gravity. If the interferometer is tilted with respect to gravity by an angle $\theta$ a dispersive phase shift of
\begin{align}
\label{phig}
\phi_{\textrm{grav}}(v)=\frac{2\pi L_g^2 }{d_g v^2}g \sin\theta
\end{align}
will result. This phase shift must be added to the total phase shift and the reference phase in the same way as the Sagnac phase. We estimate that $\theta<0.1\textrm{ mrad}$ and the corresponding uncertainty in the polarizability is less than 0.01\%.

\begin{table}
\caption{\label{ErrorTable}Systematic error budget for a single sodium measurement. The potassium and rubidium systematic error budgets are similar.}
\begin{center}
\begin{tabular}{l c c c}
\hline\hline
Source & Value(Unc.) & \% err. in $\alpha$ \\
\hline
1G-detector distance $z_{\textrm{det}}$ & 2372.4(5.1) mm & 0.43 \\
Velocity (beam shape model) & 3023(4) m/s & 0.25 \\
Detector displacement $x_1$ & 135.00(3) $\mu$m & 0.05 \\
Detector translation $\parallel$ & 50 mrad & 0.30 \\
Velocity distribution $\sigma_v$ & 149(14) m/s & 0.20 \\
$\delta_v/v$ of interfering atoms & 0.20(5)\% & 0.10 \\
Spacer thickness $a$ & 1.998(2) mm & 0.20 \\
Electrode diameter $2R$ & 12.663(25) mm & 0.10 \\
Electrode voltage $V$ & 10670(16) V & 0.30 \\
Electrode orientation ($x,y,z$) & (20,0.1,20) mrad & 0.05 \\
1G--int. region distance $z_\textrm{int}$ & 802.6(2.0) mm & 0.25 \\
Grating period $d_g$ & 100.0(1) nm & 0.10 \\
Molecule fraction & 0(1)\% & 0.10\\
Grating tilt and $g$ & 0.0(1) mrad & 0.01 \\
Beam thickness (phase avg) & 80(20) $\mu$m & 0.02\\
\\
Total Systematic Error & & 0.80 \\
\hline\hline
\end{tabular}
\end{center}
\end{table}%

\section{Conclusions and Outlook}

We measured both the absolute and relative polarizabilities of sodium, potassium, and rubidium using an atom interferometer with an electric field gradient. Furthermore, we used our ratio measurements and the more precise Ekstrom \etal measurement of sodium polarizability \cite{Eks95} to report higher precision measurements of potassium and rubidium polarizability. These measurements provide benchmark tests of atomic theories.
 
We are upgrading our apparatus to produce and detect beams of alkaline-earth atoms. We are also investigating new interaction region geometries and new ways to measure the flow velocity and velocity distribution of the atoms detected in the interferometer.  We are also using diffraction from a nanograting to study ratios of van der Waals potentials for sodium, potassium, and rubidium \cite{Lon10}.

This work is supported by NSF Grant No. 0653623. WFH and VPAL thank the Arizona TRIF for additional support.

\bibliographystyle{apsper.bst}
\bibliography{MasterBibliography7}
\end{document}